\documentstyle[psfig]{mn}

\newif\ifAMStwofonts

\ifoldfss
  \ifCUPmtlplainloaded \else
    \NewTextAlphabet{textbfit} {cmbxti10} {}
    \NewTextAlphabet{textbfss} {cmssbx10} {}
    \NewMathAlphabet{mathbfit} {cmbxti10} {} 
    \NewMathAlphabet{mathbfss} {cmssbx10} {} 
  \fi
  \ifAMStwofonts
    \ifCUPmtlplainloaded \else
      \NewSymbolFont{upmath} {eurm10}
      \NewSymbolFont{AMSa} {msam10}
      \NewMathSymbol{\upi}     {0}{upmath}{19}
      \NewMathSymbol{\umu}     {0}{upmath}{16}
      \NewMathSymbol{\upartial}{0}{upmath}{40}
      \NewMathSymbol{\leqslant}{3}{AMSa}{36}
      \NewMathSymbol{\geqslant}{3}{AMSa}{3E}

       \let\le=\leqslant
       \let\ge=\geqslant
    \fi
  \fi
\fi 

\ifnfssone
  \newmathalphabet{\mathit}
  \addtoversion{normal}{\mathit}{cmr}{m}{it}
  \addtoversion{bold}{\mathit}{cmr}{bx}{it}
  \newmathalphabet{\mathbfit} 
  \addtoversion{normal}{\mathbfit}{cmr}{bx}{it}
  \addtoversion{bold}{\mathbfit}{cmr}{bx}{it}
  \newmathalphabet{\mathbfss} 
  \addtoversion{normal}{\mathbfss}{cmss}{bx}{n}
  \addtoversion{bold}{\mathbfss}{cmss}{bx}{n}
  \ifAMStwofonts
    \ifCUPmtlplainloaded \else
      \UseAMStwoboldmath
      \makeatletter
      \new@mathgroup\upmath@group
      \define@mathgroup\mv@normal\upmath@group{eur}{m}{n}
      \define@mathgroup\mv@bold\upmath@group{eur}{b}{n}
      \edef\UPM{\hexnumber\upmath@group}
      \new@mathgroup\amsa@group
      \define@mathgroup\mv@normal\amsa@group{msa}{m}{n}
      \define@mathgroup\mv@bold\amsa@group{msa}{m}{n}
      \edef\AMSa{\hexnumber\amsa@group}
      \makeatother
      \mathchardef\upi="0\UPM19
      \mathchardef\umu="0\UPM16
      \mathchardef\upartial="0\UPM40
      \mathchardef\leqslant="3\AMSa36
      \mathchardef\geqslant="3\AMSa3E

       \let\le=\leqslant
       \let\ge=\geqslant
    \fi
  \fi
\fi 

\ifnfsstwo
  \DeclareMathAlphabet{\mathbfit}{OT1}{cmr}{bx}{it}
  \SetMathAlphabet\mathbfit{bold}{OT1}{cmr}{bx}{it}
  \DeclareMathAlphabet{\mathbfss}{OT1}{cmss}{bx}{n}
  \SetMathAlphabet\mathbfss{bold}{OT1}{cmss}{bx}{n}
  \ifAMStwofonts
    \ifCUPmtlplainloaded \else
      \DeclareSymbolFont{UPM}{U}{eur}{m}{n}
      \SetSymbolFont{UPM}{bold}{U}{eur}{b}{n}
      \DeclareSymbolFont{AMSa}{U}{msa}{m}{n}
      \DeclareMathSymbol{\upi}{0}{UPM}{"19}
      \DeclareMathSymbol{\umu}{0}{UPM}{"16}
      \DeclareMathSymbol{\upartial}{0}{UPM}{"40}
      \DeclareMathSymbol{\leqslant}{3}{AMSa}{"36}
      \DeclareMathSymbol{\geqslant}{3}{AMSa}{"3E}

       \let\le=\leqslant
       \let\ge=\geqslant
    \fi
  \fi
\fi 

\ifCUPmtlplainloaded \else
  \ifAMStwofonts \else 
    \def\upi{\pi}
    \def\umu{\mu}
    \def\upartial{\partial}
  \fi
\fi

\title[Clumpy dust disks]{Effects of Clumping on the Observed 
   Properties of  Dusty Galaxies}
 
\author[S. Bianchi et al.]{S. Bianchi$^1$\thanks{Email address:
Simone.Bianchi@astro.cf.ac.uk},
			      A. Ferrara$^2$, 
			      J.~I. Davies$^1$,
			      P.~B. Alton$^1$\\
$^1$Department of Physics and Astronomy, Cardiff University,
P.O. Box 913, Cardiff, CF2 3YB, UK\\
$^2$Osservatorio Astrofisico di Arcetri, Largo E. Fermi 5, 
50125 Firenze, Italy }

\begin{document} 

\maketitle 
 
\begin{abstract} 
We present Monte Carlo radiative transfer simulations for  spiral
galaxies modelled as a stellar disk and a two-phase clumpy dust distribution.
We divide the volume occupied by the dust into a three-dimensional grid and
assign each cell a clump or smooth medium status. Cell dimension,
clump dust mass and spatial distribution are derived from the
observed properties of Giant Molecular Clouds and molecular gas
in the Galaxy. We produce models for several values of the
optical depth and fraction of the interstellar medium residing in
clumps.  As a general result, clumpy models are less
opaque than the corresponding homogeneous models. For the adopted
parameters, the increase in the fraction of energy that escapes the disk
is moderate, resulting in surface brightness profiles that are less
than one magnitude brighter than those of the homogeneous models.
The effects of clumping are larger for edge-on views of the disk.
This is in contrast with previous preliminary results for clumping in
the literature.  We show how differences arise from the different 
parametrisation and clump distribution adopted.
We also consider models in which a fraction of the stellar radiation
is emitted within the clumps. In this case, galaxies
are less transparent than in the case when only dust is clumped.
The opacity can be even higher than in the homogeneous case,
depending on the fraction of embedded stellar emission.
We point out the implications of the results for the determination of
the opacity and dust mass of spiral galaxies.
\end{abstract} 

 
%
 
\section{Introduction}
An understanding of the effects of dust extinction in spiral galaxies
is crucial, both for deriving the intrinsic properties of the galactic
radiation field and interpreting observations of the distant universe
in the background.

While in earlier studies the opacity of a galactic disk was often treated 
in a simplistic way, causing misinterpretations and mutually exclusive
results~\cite{di89}, a number of more realistic 
models has been developed in the last years; the radiative transfer 
have been modelled for appropriate galactic geometries, including multiple
scattering, via analytical approximations~\cite{by94,si98} or Monte Carlo 
techniques (Bianchi, Ferrara \& Giovanardi 1996, hereafter BFG; De Jong 
1996; Wood 1997).
Recently, the BFG code has been exploited to produce data for a conspicuous 
set of models (Ferrara et al. 1999, hereafter The Atlas).

The majority of the radiative transfer models for spiral galaxies deal with 
smooth distributions of dust and stars. Instead, the interstellar medium is 
observed to have a complex structure. It is difficult to solve the
radiative transfer problem for a clumpy dust distribution. Therefore
most models have been implemented for simple geometries.
Analytical solutions for  a clumpy plane parallel dust layer have been 
provided by Boiss\'e \shortcite{bo90} for a two-phase medium and 
by Hobson \& Scheuer \shortcite{ho93} for a generic N-phase medium, under the
assumption of isotropic scattering.
Witt \& Gordon \shortcite{wi96} and Wolf, Fischer \& Pfau \shortcite{wo98}
performed Monte Carlo simulations for the radiative transfer through a 
spherical, 
two-phase dusty medium, illuminated by a central source. They divided 
the sphere in a cubic lattice, assigning each cell a low or high density
status according to chosen values for the filling factor and density
ratio of smooth/clump medium. Spherical clumps inside a spherical dust
distribution have been studied by V\'arosi \& Dwek \shortcite{va99}, allowing
radiation to come from a central point source, a uniform
distribution of emitters, or a uniform distribution of external sources;
analytical approximations are tested against Monte Carlo simulations.

We summarise here a few of the properties derived for clumpy distributions
of dust. As a general rule, a clumpy medium has a higher transparency 
with respect to a homogeneous one with the same mass of dust.
The effect of
scattering is reduced, because light entering a dense clump has a low
probability of escaping: from the point of view of the radiative transfer,
a dense clump therefore behaves similarly to a dust grain with smaller
albedo than the one characterising single grains in a smooth medium.  
For a clumpy distribution the amount of energy absorbed varies less with
the wavelength of radiation than for a homogeneous distribution: this 
concurs, together with  geometric effects (see The Atlas), in an 
apparent extinction, or more appropriately attenuation, curve flatter than the 
actual one.

However, the details of radiative transfer in a clumpy medium depend  
not only on the parameters used for the clumping itself, but also on the 
distribution of dust with respect to the stars. 
In this paper we present a study of the
radiative transfer through a clumpy dust distribution for geometries 
typical of a spiral galaxy. Although the ISM may have a wide range of
densities \cite{ha94,he98}, we adopt a simple two-phase scheme for
the dust distribution: a smooth medium associated with the low-density 
diffuse atomic gas and clumps associated to high density Giant Molecular 
Clouds (GMCs). The distribution of clumps is derived from observations 
of the molecular gas in the Galaxy. 

Only a few works include clumping in a radiative transfer model
appropriate for galactic geometries. Kuchinski et al.~\shortcite{ku98} use
preliminary results from a Monte Carlo model (based on Witt \& Gordon
1996 formalism) to derive opacities of
edge-on galaxies from their colour gradients. Although a few aspects of
the inclusion of clumping are presented in that paper, the authors defer a 
detailed discussion to a forthcoming paper. Clumping has been also included by
Sauty, Gerin \& Casoli \shortcite{sa98} in their model of the FIR and C$^+$
emission from the spiral galaxy NGC~6946; however, it is difficult  to
single-out the effect of clumping from the results shown.

Since star formation occurs in the dense phase of the
interstellar medium, it is justified to assume that a fraction of the
stellar emission occurs within clumps of dust. This may change the
effects of clumping. In their model for a centrally illuminated dust
distribution, V\'arosi \& Dwek \shortcite{va97} find that the absorption 
efficiency may
increase, rather than decrease, when clumps have a fractal distribution:
being a fractal distribution a more connected set with respect to
a uniform random distribution of spherical clumps, the fractal cloud 
can behave like a shield in front of the source. Embedded stars are
included in the radiative transfer model of Sauty et al.~\shortcite{sa98} and
Silva et al.~\shortcite{si98}. In this paper we will also consider the effects 
of a clumpy distribution for the radiation sources.

The paper is organised as follows. In \S~2 we present a
detailed description of the parameters used to describe the stellar
emission and the clumpy dust distribution, referring to BFG and The Atlas 
for the general treatment of the radiative transfer; \S~3 is dedicated
to the effect of clumping for radiation coming from a smooth stellar
distribution; \S~4 shows the results when the possibility of  sources
embedded in dust clumps is taken into account. Finally, \S~5 contains a
summary and a discussion of the application of the results obtained in
this work.  Throughout the rest of this paper we will use the notation 
CM to refer to models including clumping and HM for homogeneous models.

\begin{table*}
\begin{minipage}{11cm}
\caption{Parameters of the clumpy distributions.}
\label{tabmodel} 
\begin{tabular}{rccrccrccrcr}
&$f_c$&\multicolumn{2}{c}{0.25}
     &&\multicolumn{2}{c}{0.50}
     &&\multicolumn{2}{c}{0.75}  && M$_{dust}$ \\
$\tau_V$&&$\tau_V^s$& N$_c$&&$\tau_V^s$& N$_c$&&$\tau_V^s$& N$_c$&&10$^{6}$M$_\odot$\\ \hline
0.1 && 0.075 & 255  && 0.05 & 511   && 0.025 & 784 &&0.7  \\ 
0.5 && 0.375 & 1277 && 0.25 & 2554  && 0.125 & 3831&&3.4  \\ 
1.0 && 0.75  & 2554 && 0.5  & 5108  && 0.25  & 7662&&6.8  \\ 
2.0 && 1.5   & 5108 && 1.0  & 10215 && 0.5   &15476&&13.6  \\ 
5.0 && 3.75  &12769 && 2.5  & 25539 && 1.25  &38308&&34.0  \\ 
10.0&& 7.5   &25539 && 5.0  & 51077 && 2.5   &76616&&68.0  \\ \hline
\end{tabular}

\medskip
For each value of the
fraction of gas in clumps $f_c$ and of the optical depth $\tau_V$, the
optical depth of the smooth medium $\tau_V^s$ and the total number of
clumps $N_c$ are listed. The final column gives the total dust mass of
the model.
\end{minipage}
\end{table*}

\section{The Model}

The results are obtained using an adapted version of the 
Monte Carlo code for the radiative transfer in dusty galaxies described in BFG.
The original code has been simplified: a single Henyey-Greenstein scattering 
phase function (Henyey \& Greenstein 1941; see also BFG) is used,
with empirically derived values for the albedo
$\omega$ and of the asymmetry parameter $g$. We use here the values
$\omega=0.6$ and $g=0.6$\footnote{The values for $\omega$ and $g$ are
measured in Galactic reflection nebulae, by comparing the intensity of
scattered light with radiative transfer models. Since a clumpy medium
has a lower effective albedo, neglecting clumping in the model
may bias the measure of $\omega$ towards lower values, especially for
optically thick clouds and stronger forward scattering \cite{wi96}.}, 
appropriate for radiation in the V-band \cite{go97}.
The polarisation part of the radiative transfer code has been omitted. 
The same approach has been used for The Atlas.

In this paper we restrict ourselves to a single stellar spatial 
distribution, a three-dimensional disk with exponential behaviour
in both horizontal and vertical directions. The horizontal and 
vertical scale lengths have been chosen to match the observed values 
for the old disk population of the Galaxy, $\alpha_\star = 3$~kpc 
\cite{ke91,fu94} and $\beta_\star =
0.26$~kpc, respectively. Note that in BFG and in The Atlas we used 
$\alpha_\star =4 $~kpc \cite{ba90}: while in the previous models
only the ratio between the scale lengths was relevant, for the model of
this paper the absolute values must be specified, since the physical 
dimensions of the clumps are required (see later).
The ratio between the horizontal and vertical scale lengths used
in this paper is the same as in BFG and the Atlas.

Some specified fraction of the total dust mass is distributed in 
a smooth exponential disk, similar to the stellar one, with scale lengths 
$\alpha_d = \alpha_\star$ and $\beta_d = 0.4 \beta_\star $ (BFG and Atlas), 
while the remaining mass is distributed in clumps. 

The total mass of dust (smooth medium + clumps) is defined in terms of 
the optical depth of the model $\tau_V$, i.e. the V-band face-on optical
depth through the centre of the disk for a HM. For the
exponential disk, the mass is easily computed using the formula
\begin{eqnarray*}
M_{\mathrm{dust}}&=&\tau_V\;\alpha_d^2\;\frac{8\pi}{3}
\frac{a\delta}{Q_V}\;\left[1-(n+1)e^{-n}\right]\\
&=&6.8\times 10^6 \;\tau_V \;\;\;\left[\mbox{M$_\odot$}\right],
\end{eqnarray*}
where $a=0.1\mu$m and $\delta=3\mbox{g cm$^{-3}$}$ are the grain 
radius and density, $Q_V=1.5$ the extinction efficiency in the V-band 
\cite{hi83} and $n=4.6$ is the radial truncation of the 
disk (in scale length units; see later). In the following, when a CM
is said to have been produced for a value of $\tau_V$, it means that
the CM has the same dust mass as a HM with a face-on V-band optical 
depth $\tau_V$. Therefore, only in the case of a HM $\tau_V$ is a 
real optical depth, while for CMs is mainly an indicator of the mass
of dust in the model, the effective opacity depending in a complex way
on  $\tau_V$ and the other parameters for the clumpy dust distribution.
Models have been produced for optical depths $\tau_V=$ 0.1,
0.5, 1, 2, 5, 10.

Using a Galactic gas-to-dust ratio of 150 \cite{de90,so94}, the gas mass 
can be derived from the dust mass. A fraction $f_c$ of the total gas mass 
is then attributed to the molecular component. In this paper we explore
three different values, $f_c=$ 0.25, 0.5, 0.75. These values are
representative of actual ratios observed in late type galaxies
(from Scd to Sb, respectively; Young \& Scoville 1991). 
All of the molecular gas (and the associated mass of dust) is
supposed to be distributed in clumps. Our choice is appropriate,
most of the molecular gas ($\approx$ 90\% in mass) in the Galaxy being
in the form of GMCs \cite{co91}.

As for the clumpy structure, we use the same two-phase formalism
as in Witt \& Gordon \shortcite{wi96}.
The space occupied by dust is divided into cubic cells of the 
dimension of Galactic GMCs. Blitz \shortcite{bl91} quotes a
typical diameter of 45 pc: a cubic cell of 36 pc will therefore have 
the same volume as a molecular cloud. 

First, each cell is assigned a value corresponding to the local 
absorption coefficient (i.e. the inverse of the light mean free path
through dust) for the smooth distribution of dust.
If $\tau_V$ is the optical depth of the model, a fraction 
(1-$f_c$) of the dust mass is distributed in the smooth medium.
Thus, the smooth medium opacity is defined by a face-on optical
depth $\tau^s_V= \tau_V*(1.-f_c)$. The local absorption coefficient for
the smooth medium is then computed from the assumed double-exponential
distribution, using $\tau^s_V$.

Second, we distribute the clumps. The position of each clump is derived 
from the distribution of molecular gas in the Galaxy, using the Monte
Carlo method. 
We have used the CO observations of the first Galactic quadrant described 
by Clemens, Saunders \& Scoville \shortcite{cl88}.
The radial density distribution has been derived from the plot of the
mass of molecular hydrogen vs. galactocentric radius in their Fig. 11;
we have adopted a gaussian vertical distribution, as parameterised in
their Eqn.~3, with a FWHM that increases with galactocentric
distance. Following the molecular gas, the distribution of clumps is more
concentrated on the Galactic plane than the smooth dust. At any 
galactocentric distance, the probability of finding clumps for $|z|>2.5
\beta_d$ is nil.
The distribution has been rescaled assuming 8.5 kpc as
the Sun distance from the Galactic centre: the position of the Galactic
molecular ring is thus between 1.5 and 3.5 radial scale lengths from 
the centre. Outside 13 kpc ($\approx$ 4.5 $\alpha_\star$) the molecular
gas is not detected (see also Heyer et al.~1998). In the model, 
both the dust and the stellar disks are truncated at this distance from the
centre.

Each clump has been given a typical GMC mass of $10^{5}$ 
M$_\odot$ \cite{blXX}. The total number of clouds in 
the model is then directly derived from the mass of gas in clumps. 
Assuming that the dust has a uniform density inside each cloud, the 
absorption coefficient in the clump
can be easily computed from the clump gas mass and 
dimensions, using the adopted dust properties and gas-to-dust ratio:
each clump has an absorption coefficient in the V-band of 110 kpc$^{-1}$,
corresponding to an optical depth $\tau_{V}\approx 4$.
Since the effects of a clumpy structure are stronger when individual clumps 
are optically thick \cite{bo90,bo96} our
simulations provide an upper limit to the effects of clumping in a 
spiral galaxy.

For all the cells that have been assigned a clump status, the clump
absorption coefficient is summed to the absorption coefficient of the
smooth medium. The optical depth along any photon path is computed
integrating the absorption coefficient array along the specified
direction. Apart from this, the Monte Carlo code follows
the same scheme as described in BFG. A list of the parameters defining
the clumpy dust distribution is given in Table~\ref{tabmodel}.

To reduce the computational time and the dimension of the array involved, 
we use only one octant of the galaxy and assign values in the other
octant according to the model symmetries: the dust absorption
coefficient is stored for 383x383x12 cells. Because of this, clumps 
are distributed randomly only in one octant. This does not affect the
statistical properties of the model, as long as the number of clumps is
large and the position of emission of photons is random. Nevertheless, we
have tested the validity of this simplification comparing the results
to a case with clumps distributed randomly all over the space, obtaining
the same results.
Finally, simulated images of 201x201 pixels are produced for several
inclinations (BFG). 

In the following sections, we present results and plots for a few
representative cases of optical depth and inclination. The whole 
dataset of this paper is available at
http://www.arcetri.astro.it/\~~sbianchi/clumping.html

\section{Results}
\begin{table*}
\begin{minipage}{13cm}
\caption{Fraction of energy absorbed in each model.}
\label{tabfrac}
\begin{tabular}{rccccccccccccccc} 
&$f_{emb}$&     &&\multicolumn{3}{c}{0.00}
&~&\multicolumn{3}{c}{0.15}
&~&\multicolumn{3}{c}{0.50} \\
& $f_c$
     &0.00 &&0.25 &0.50 &0.75 &&0.25 &0.50 &0.75 &&0.25 &0.50 &0.75\\
$\tau_V$& &&&&&&&&&&&&&\\ \hline
 0.1&&0.02 &&0.01 &0.01 &0.01 &&0.09 &0.08 &0.08 &&0.25 &0.25 &0.25\\
 0.5&&0.07 &&0.06 &0.05 &0.04 &&0.13 &0.12 &0.11 &&0.29 &0.28 &0.28\\
 1.0&&0.12 &&0.11 &0.09 &0.07 &&0.17 &0.16 &0.15 &&0.32 &0.31 &0.30\\
 2.0&&0.19 &&0.18 &0.15 &0.12 &&0.24 &0.22 &0.19 &&0.38 &0.36 &0.35\\
 5.0&&0.32 &&0.31 &0.28 &0.23 &&0.37 &0.34 &0.30 &&0.49 &0.47 &0.44\\
10.0&&0.44 &&0.43 &0.39 &0.34 &&0.48 &0.45 &0.40 &&0.59 &0.57 &0.54\\
\hline
\end{tabular}
{For each value of
$\tau_V$, the first column gives the fraction for the HM 
($f_c=0$), then each group of three columns gives the values for
CMs with a specific fraction of the total energy emitted within
clumps ($f_{emb}$; $f_{emb}=0$ refers to the case with a homogeneous
stellar distribution) and for the three values of the fraction of gas in
clumps $f_c$.}
\end{minipage}
\end{table*}

As already described in the introduction, the main effect of clumping is 
that of an overall increase in the transparency of the model. 
In Table~\ref{tabfrac} we show the fraction of the total energy that is
absorbed in each model, as a function of $\tau_V$ (that defines the mass
of dust) and $f_c$. Obviously, for the same value of the optical depth,
more energy is absorbed in a HM (defined in the
table by $f_c$=0.0). The results presented for the HM are
the same as for the spiral galaxy disk in The Atlas. When the clumping is
introduced, CMs  with larger  $f_c$ are 
increasingly more transparent. 

\begin{figure}
\centerline{\psfig{figure=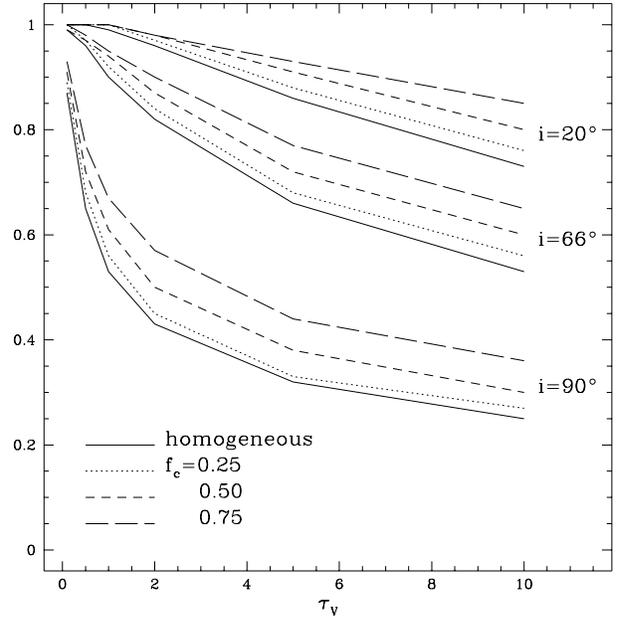,width=9cm}}
\caption{Attenuation as a function of the optical depth
$\tau_V$, for the HM and CMs with
$f_c$=0.25, 0.50, 0.75. Data are plotted for
three inclinations, $i$=20$^\circ$, 66$^\circ$ and 90$^\circ$.  
}
\label{atte_tau}
\end{figure}

Within the same model, the effects of clumping depend also on the
disk inclination.
This is clearly shown in Fig.~\ref{atte_tau},
where we have plotted the attenuation (i.e. the ratio between the
observed flux and the intrinsic, unextinguished, total one)
as a function of the optical depth, for the HMs and the CMs
for three values of the inclination angles,
$i$=20$^\circ$, 66$^\circ$, 90$^\circ$.
As seen before, CMs have the same attenuation 
as HMs with a smaller $\tau_V$, for any inclination. 
The largest differences
between the two cases are found in more face-on cases, and for
high values of $f_c$. For instance, a face-on CM
$f_c$=0.75 and optical depth $\tau_V=10$ has the same attenuation as an
HM with $\tau_V\approx 4$, while for an inclination 
of 20$^\circ$, it corresponds to a HM with 
$\tau_V\approx 5$. For the same values of  $\tau_V$ and $f_c$, the
smooth medium has an optical depth  $\tau_V^s$=2.5. 

\begin{figure*}
\centerline{\psfig{figure=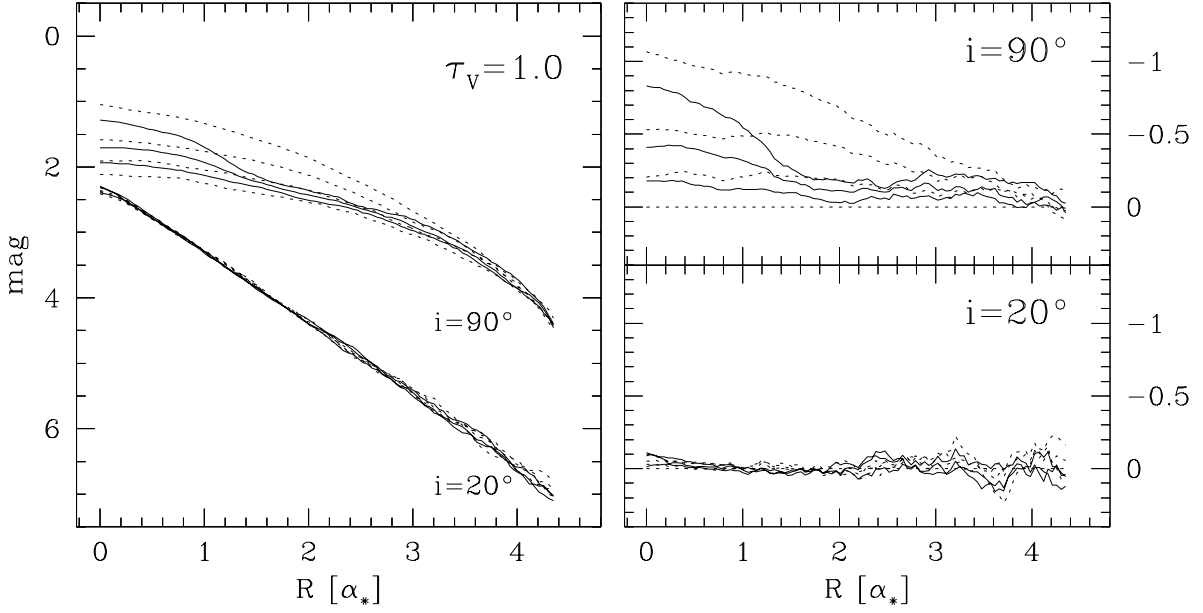,width=17cm}}
\caption{Major axis profiles (left panel) for models with $\tau_V$= 1.0 
and inclination $i$=20$^\circ$ (face-on), 90$^\circ$ (edge-on). 
Solid lines refer to CMs, for $f_c$= 0.25,
0.5, 0.75. The brighter profiles corresponds to the case with higher 
fraction of gas in clumps, $f_c$= 0.75, while the dimmer to the case
with $f_c$= 0.25. HMs are presented for
comparison (dotted lines). The brighter profile corresponds to a HM
with the same optical depth as the homogeneous dust ($\tau_V^s$)
for the case
$f_c$=0.75, followed by the analogous profiles for $f_c$= 0.5, 1.0.
The dimmer profile refers to a HM with the optical
depth $\tau_V$. 
The right panel presents the difference in magnitude between each model
and the HM with the optical depth $\tau_V$.
All curves have been smoothed with a box of 10 pixels.
}
\label{thinpro}
\end{figure*}

\begin{figure*}
\centerline{\psfig{figure=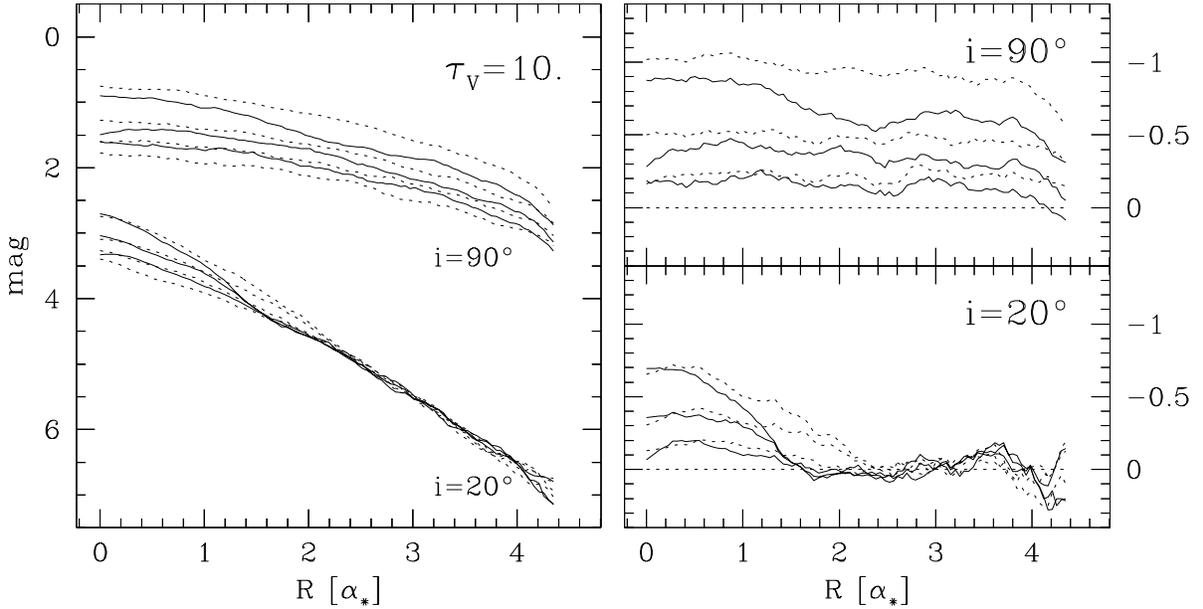,width=17cm}}
\caption{Same as Fig.~\ref{thinpro}, but for models with $\tau_V$=
10.0. 
To avoid overlap, a constant value of 2 magnitudes has been
subtracted from the surface brightness for the edge-on major
axis profile presented in the left panel.
}
\label{thickpro}
\end{figure*}

\begin{figure*}
\centerline{\psfig{figure=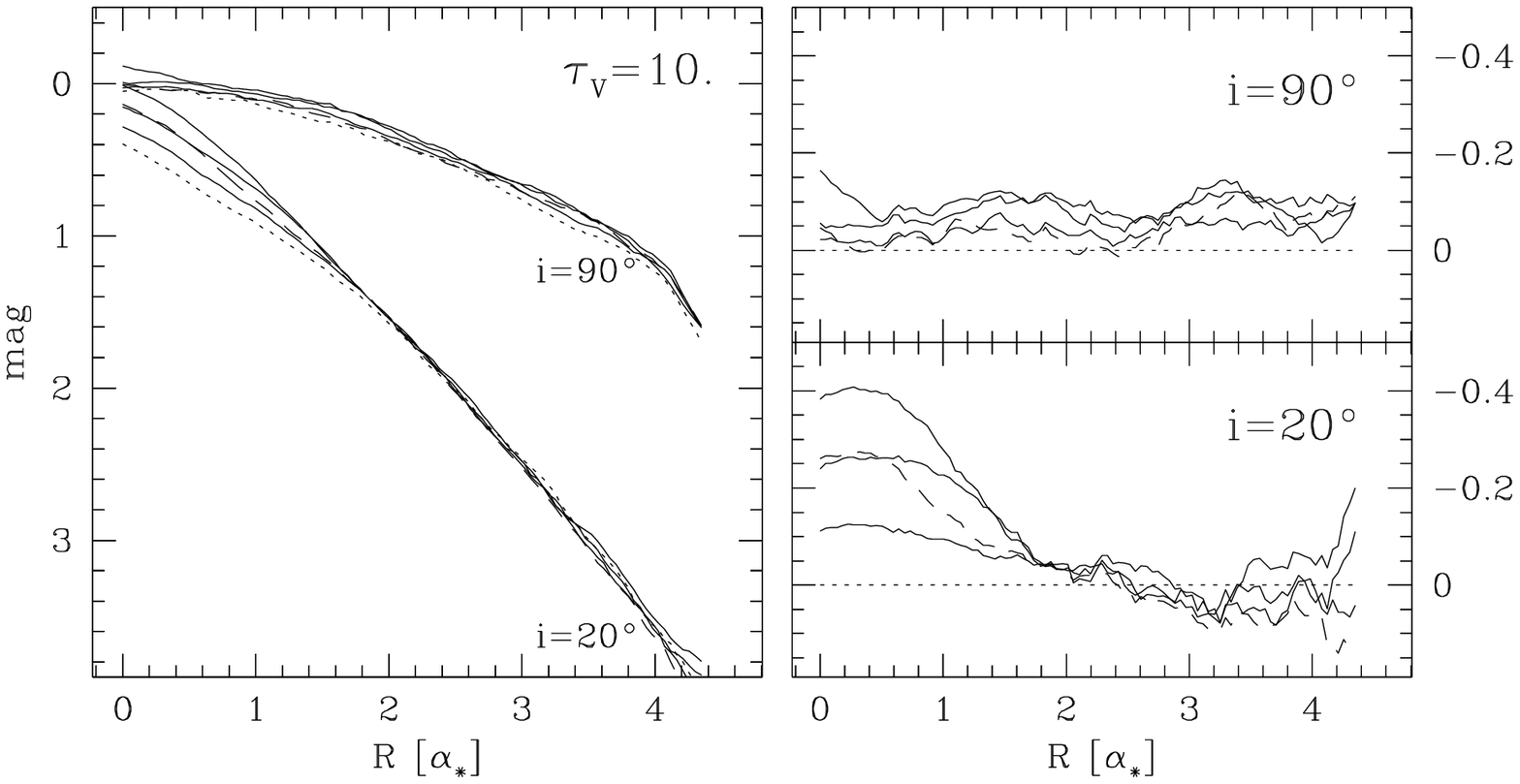,width=17cm}}
\caption{Major axis profiles (left panel) for CMs with constant filling 
factor all over the dust distribution and a ratio of 100 between densities 
in clumps and in the nearby smooth medium (See text for details). 
Models have $\tau_V$=10 and $i$=20$^\circ$ and 90$^\circ$. Solid lines 
refer to $f_c$= 0.25, 0.5, 1.0 (Brighter profile). With a dashed line the
case $f_c$= 0.95 is plotted, while the dotted line refers to the
HM with the same $\tau_V$. 
To avoid overlaps, a constant value of 0.75 magnitudes has been
subtracted from the surface brightness for the edge-on major
axis profile.
The right panel presents the difference in magnitude between each model
and the HM with the optical depth $\tau_V$.
All curves have been smoothed with a box of 10 pixels.}
\label{confro}
\end{figure*}

The different behaviour of CMs with inclination can also be seen analysing 
the disk major axes profiles. In
Fig.~\ref{thinpro} and \ref{thickpro} we show the major axis profiles
for two representative cases of low ($\tau_V=1$) and high ($\tau_V=10$)
optical depth, for inclinations of $i$=20$^\circ$, and 90$^\circ$.
In each left panel we plot with solid lines the profiles for CMs: 
the brighter profile always refers to the case with higher
fraction of dust in clumps, $f_c=0.75$, while the fainter is for $f_c=0.25$.
As a comparison, we have also plotted a series of profiles for HMs
(dotted lines); the three brighter profiles have the same
optical depth $\tau_V^s$ as the smooth medium in each of the CM, 
from $f_c=0.75$ (upper profile) to $f_c=0.25$. The faintest of
the HM is characterised by the optical depth $\tau_V$. 
The differences between each profile and the profile for the HM 
with optical depth $\tau_V$ are shown in the right panels, for the
chosen inclinations.

As expected, a CM characterised by a value of 
$\tau_V$ and $f_c$ has a major axis profile at an intermediate
brightness between the HM of optical depth $\tau_V$
and the HM of optical depth $\tau_V^s$, i.e.\ the homogeneous smooth 
component.  The differences between HM 
and CM are not large, and always smaller than 1 mag.
For optically thin cases, models are virtually indistinguishable at low 
inclinations (as in the $\tau_V=1$ profiles of Fig.~\ref{thinpro}).
For optically thick cases (e.g. the models in Fig.~\ref{thickpro}), the 
profiles for CM are closer to those for the smooth medium only for
$R< \alpha_\star$, where the number of clumps, and therefore their
filling factor, is small. Between 1 and 3 $\alpha_\star$,  where
the molecular distribution peaks (the Galactic 
ring), the profile is closer to that for the HM 
with the same optical depth.
On the contrary, profiles for edge-on cases are always brighter than 
the corresponding HM.

These results are in contrast with those presented 
by Kuchinski et al.~\shortcite{ku98}. They derived opacities of highly inclined
galaxies, by comparing optical and near-infrared colour gradients with Monte 
Carlo radiative transfer simulations, including scattering and clumpy 
dust distribution.  They found that the derived optical depths are 
insensitive to the dust having either homogeneous
or clumpy distribution. This is explained by the significant number of clumps 
intersected by any line of sight through  a nearly edge-on galaxy.
On this basis they argue that clumping effects may be more important for lower 
inclinations, where some lines of sight pass across clumps, while 
others do not. However, the description of the clumpy structure of dust
in Kuchinski et al.~\shortcite{ku98} is different from the one adopted here.
Following Witt \& Gordon \shortcite{wi96}, they assign a constant filling factor
($ff$=0.15) for the high density cells all over the galaxy, and assume
a ratio of 100 between densities in clumps and in the nearby smooth 
medium. Since the smooth medium has an exponential distribution, clumps
close to the galactic centre have a higher optical depth than those
in the outer disk. Clumps in our simulations, instead, are distributed
into a ring-like structure and have the same optical depth. We feel these
assumptions are more realistic and are supported by the available 
observational data.

To clarify the reasons for the discrepancy, we have modified the code to deal
with the Kuchinski et al.~\shortcite{ku98} formalism. Clumps are now 
characterised
by a constant filling factor all over the dust distribution and by a
value for the local ratio of densities of the two dust phases. For these
tests we have used the value 100 for the density ratio. The geometry of
the galaxy and the cells dimensions have been kept as in \S 2.
In Fig.~\ref{confro} we show the major axis profiles in the case	
$\tau_V$=10, for the two inclination $i=20^\circ$ and $90^\circ$.
The solid lines refer to CMs with the usual values $f_c$=0.25,
0.50, 0.75. In these models $f_c$ can be directly converted into the
filling factor $ff$ using the formula
\[
ff=\frac{f_c}{f_c+k(1-f_c)},
\]
where $k=100$ is the density ratio.
The $f_c$ values of this paper thus correspond to quite low global filling 
factors, 
$ff$=0.003, 0.01, 0.03, respectively. The dashed line is for $ff$=0.15, as in
Kuchinski et al.~\shortcite{ku98}. For the assumed density ratio, this value of
the filling factor corresponds to $f_c\approx 0.95$, i.e. 95\% of the 
total mass of dust in the model is distributed in clumps.
Indeed, for this simulation setup, CMs
have profiles closer to the HM when viewed at larger 
inclinations. 

\begin{figure*}
\centerline{\psfig{figure=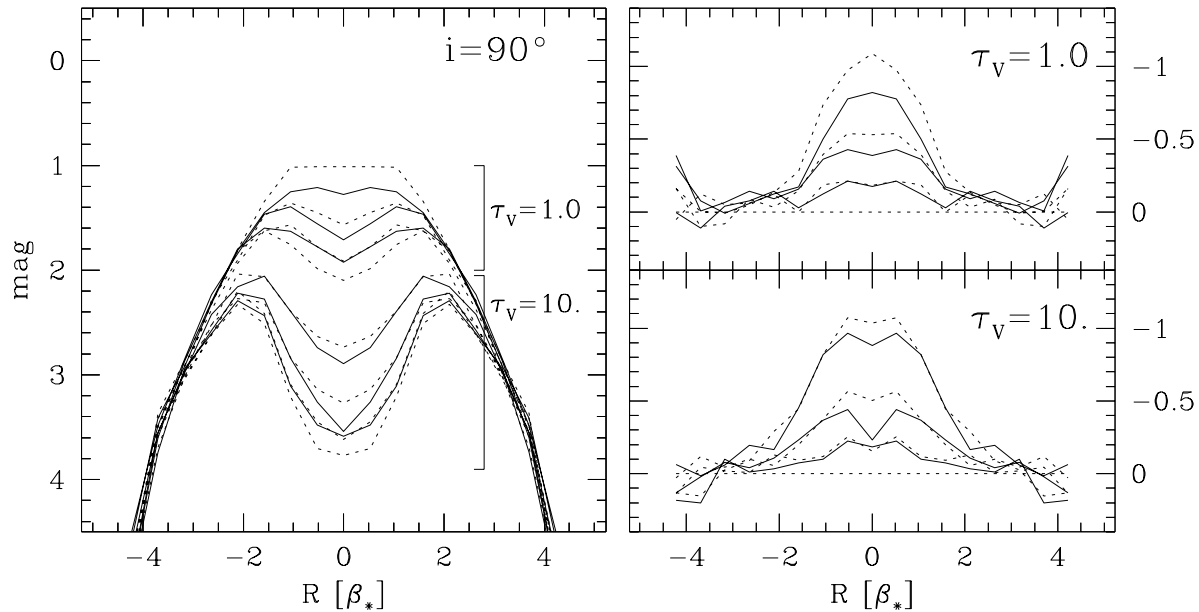,width=17cm}}
\caption{Edge-on ($i=90^\circ$) minor axis profiles (left panel) 
for models with $f_c$= 0.25, 0.5, 0.75 (solid lines).  As in 
Fig.~\ref{thinpro}, brighter profiles corresponds to higher values of 
$f_c$.  Dotted lines refers to HMs as described for Fig.~\ref{thinpro}.
The right panel presents the difference in magnitude between each model
and the HM with the optical depth $\tau_V$.
}
\label{allmin}
\end{figure*}

As for the profiles in Fig.~\ref{thinpro} and Fig.~\ref{thickpro},
an increase of $f_c$ in the range 0.25 - 0.75 produces an increase
in the global transparency of the dust distribution, that is shown 
in Fig.~\ref{confro} by the brighter profiles (especially in the face-on
case) when a larger fraction of gas is distributed in clumps. The effects 
of clumping are instead reduced for $f_c$=0.95 ($ff$=0.15).
This is consistent with the simulations of radiative transfer in a clumpy
media presented by V\'arosi \& Dwek~\shortcite{va99}. They have studied the
case of a clumpy dust sphere under different radiation fields: a central
source, a homogeneous distribution of isotropic emitters and a uniform
external field. When all the other parameters defining the dust
distribution are fixed, the fraction of absorbed photons as a function of
the clumps filling factor shows a minimum, for any of the radiation
fields. The exact value of $ff$ for which the dust distribution has a 
maximum transparency depends on the details of the model. In our
simulation, for the value of the clump/smooth medium density ratio
adopted, clumping has the major effect in the reduction of the fraction
of absorbed energy for $0.03\le ff< 0.15$ (or 0.75$\le f_c<$0.95).
In the CM for $f_c$=0.95, 43\% of the radiation is absorbed, almost
the same quantity as for the $f_c$=0.25 case (and for the HM). 
As already stated, the model for $f_c$=0.75 is more transparent,
with 38\% of the total radiation absorbed.
With respect to the main models of this paper, the constant filling factor 
distributions are slightly more opaque, for the same value of $f_c$.
In the $\tau_V$=10 model with $f_c$=0.75 of Table~\ref{tabfrac}, for 
instance, the fraction of absorbed energy is 34\%.

The differences in behaviour between the main CM of this
paper and the one with constant filling factor arise because of two
reasons: (i) the different number of clumps for a given value of $f_c$
and (ii) the different geometrical distribution. In the main CM clumps have
all the same mass, and therefore the same density, for a given cell
dimension. In the constant filling factor CM the density of a
clump depends on the local density of the smooth medium (it is 100 times
for the model analysed here). Because of the exponential distribution of 
the smooth medium, clumps at larger distance from the centre have smaller 
density (and mass) than those in the inner part of the galaxy. 
Consequently, when the fraction of ISM mass that is locked in clumps is 
fixed, the main CM will have a smaller number of clumps of higher
mass than the one with a constant filling factor. For the case 
$\tau_V$=10 with  $f_c$=0.75  analysed before there are roughly 4 times
more clumps in the latter than in the former. This smaller number of
clumps is then distributed on preferential places in the main CM,
more concentrated on the galactic plane and in the molecular ring.
When seen face-on the large number of clumps in the ring position
attenuates a lot of the radiation coming from the half of the galaxy
below the plane, acting almost as a screen. Only outside the ring the 
filling factor of clumps is small, and in this regions the face-on
profile is closer to that for a HM with the optical
depth of the smooth medium. In the edge-on case, instead, it is the
light emitted outside the ring, attenuated by the smooth dust
distribution, that dominates the radiative transfer, thus resulting in
a major axis profile of intermediate brightness between those of the
HMs with $\tau_V$ and $\tau_V^s$.

In Fig.~\ref{allmin} we plot the edge-on minor axis profiles for the
main model with optical depths $\tau_V=1$ and 10.
Geometrical and optical parameters of the dust distribution
in spiral galaxies are mainly retrieved by fitting radiative transfer
models to images of edge-on galaxies \cite{xi97,xi98,xi99}.
At this inclination, in fact, we can have separate views of the radial
and vertical dust distribution. Furthermore, dust extinction is increased 
as a result of the projection. The profiles of Fig.~\ref{allmin} show
a behaviour analogous to the edge-on major axis profiles in
Fig.~\ref{thinpro} and ~\ref{thickpro}, with a reduced extinction that
results in shallower extinction lanes on the galactic plane.
On the contrary, edge-on minor axis profiles for CMs with constant
filling factor are virtually indistinguishable from the HM.

\section{Embedded sources}

\begin{figure*}
\centerline{\psfig{figure=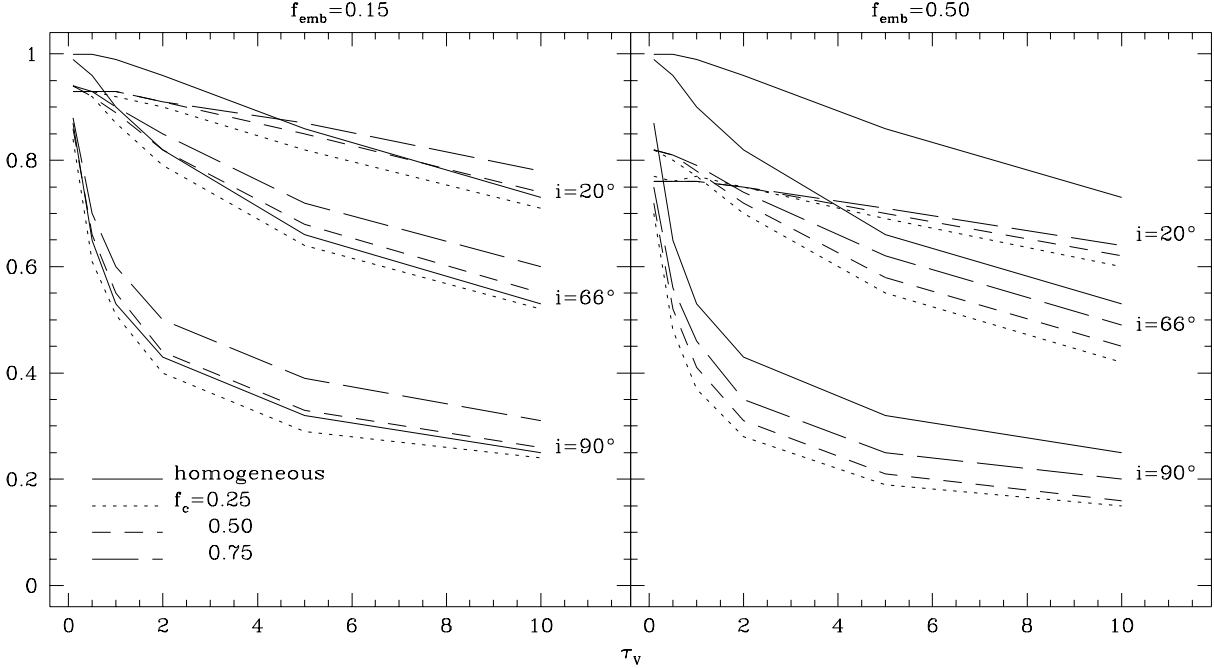,width=17cm}}
\caption{Same as Fig.~\ref{atte_tau}, but for models with $f_{emb}$=0.15 and
0.50.}
\label{atte_tau2}
\end{figure*}

So far clumping has been introduced only in the description of the
dust distribution.
Since we use a clumpy structure to simulate GMCs, it is logical to 
assume that part of the stellar emission comes directly from inside 
the clumps, the stars being formed in the higher density phases of the ISM.
In this section we study the effects of clumping both for dust and
stars.

Radiation from stars embedded in GMCs is
simulated assuming that a fraction $f_{\mathrm{emb}}$ of the total
energy emitted by stars comes from inside the clumps of dust. We have
used the values 0.15 and 0.50. The actual value of this parameter is
difficult to be derived from observations, since it would require sensitive
indicators of the (optically) obscured star formation rates. 
The above range is an educated guess that comes from the estimate of the
average fraction of stellar lifetime spent inside dark clouds. 
Wood \& Churchwell (1989) and Churchwell (1991), from their study of
UCHII regions, concluded that O stars are embedded in their natal
molecular clouds on average about 15\% of the main sequence of an O6
star. However, this might be a lower limit and the
fraction of total embedded star formation could be as high as 50\%
when the contribution from low mass stars is taken into account. 
Within the Monte Carlo code, a fraction $f_{\mathrm{emb}}$ of the
photons is emitted inside the cells. The position of the emission inside 
the cell is randomly distributed within its boundaries.
This allows more radiation to escape clumps, with respect to the case where 
the photons are emitted in the centre of the cell. Once the photon is
emitted, the radiative transfer is carried out through the clumpy dust
distribution as described in \S~2.

The fraction of absorbed energy in CMs with embedding is shown in 
Table~\ref{tabfrac}, for the two values of $f_{emb}$. 
In general, a larger fraction of energy is absorbed in these CMs
with respect to the case when only dust has a clumpy distribution. The
fraction of absorbed energy may be even larger than for a HM. 
When the degree of clumpiness increases (larger $f_c$) the
effect of embedding is mitigated and more energy can escape. As an
example, for $f_{\mathrm{emb}}$=0.15, optically thick CMs
with $f_c\ge 0.5$ have an overall opacity similar or smaller than
for the HM. The absorption in the case $f_{\mathrm{emb}}$=0.5
is larger and although it decreases with increasing $f_c$ as for the
previous case, the fraction of absorbed energy is always higher than in
the HM.
The large fraction of energy absorbed in the 
optically thin cases, especially for $f_{emb}$=0.5 might seem surprising. 
This is because in our
formalism, for optically thin cases (small mass of dust), the embedded
emission comes from relatively few opaque clumps. For a cubic cell of optical
depth 4 through its side and with a homogeneous distribution of internal
emitters, nearly 50\% of the radiation is absorbed within the cell.
For instance, in a model with $\tau_V$=0.1 and $f_{emb}$=0.5, half of the
radiation is emitted within clumps, and half of it is absorbed, thus
resulting in a fraction of absorbed energy 0.25, as in Table~\ref{tabfrac}.

The global increase in extinction can also be seen in the attenuation plots of
Fig.~\ref{atte_tau2}. Models with clumping distributions for both stars and 
dust have a transparency lower than those with a clumpy dust distribution 
only, at any inclination. As already shown before, for $f_{emb}=0.50$, all the 
CMs are more opaque than the corresponding HM with the same 
optical depth. 
For the more face-on inclinations, the behaviour of the attenuation is less
dependent on $\tau_V$ with respect to the HM. This is because the
projected optical depth of the smooth medium is lower, and the clumps dominate 
the emission (and absorption). 

\begin{figure*}
\centerline{\psfig{figure=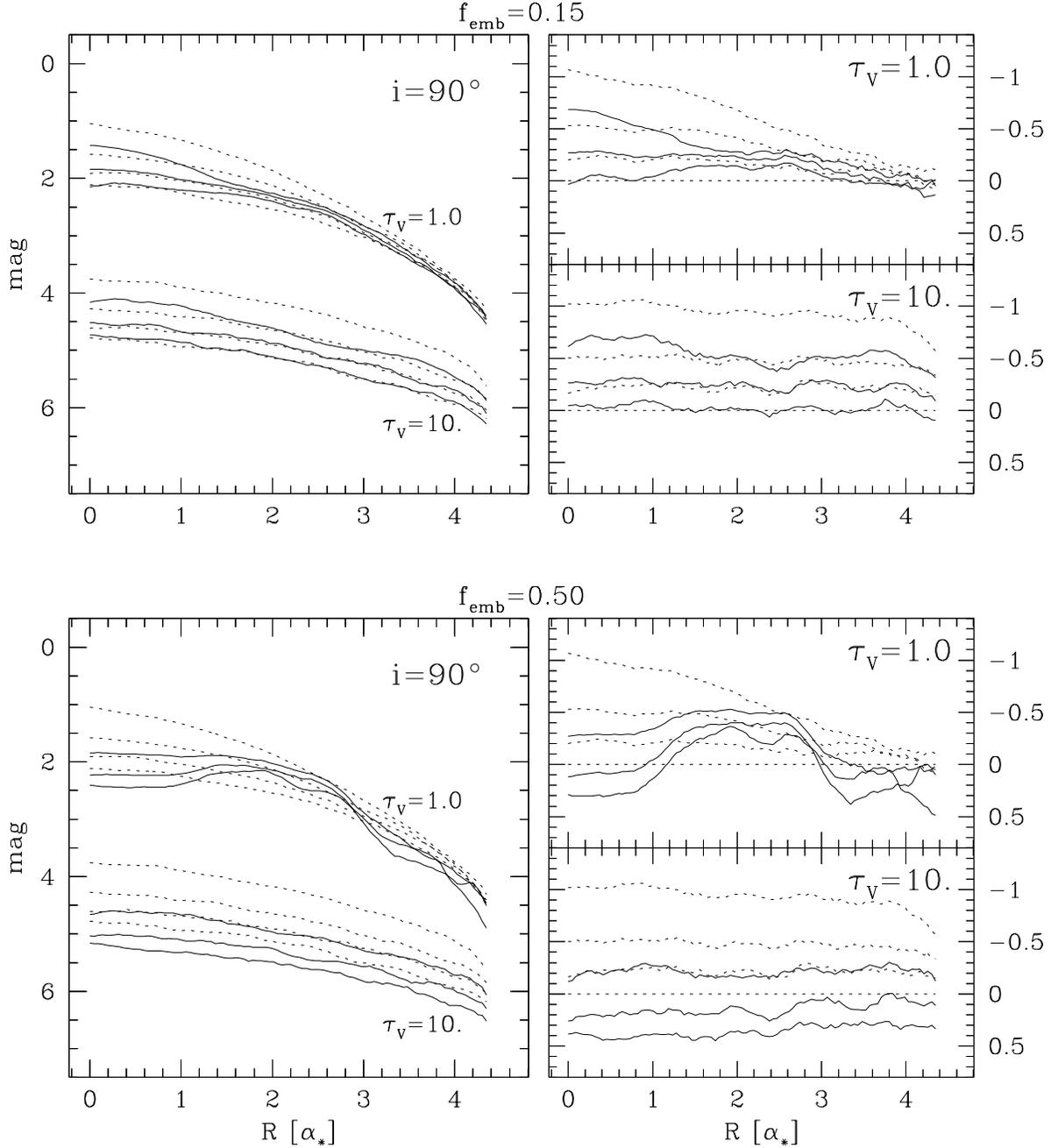,width=17cm}}
\caption{Edge-on major axis profiles (left panels) for models with embedded 
sources.  Profiles are plotted for optical depths $\tau_V$=1, 10 and for the
three values of $f_c$ (solid lines). As in Fig.~\ref{thinpro}, brighter
profiles corresponds to higher values of $f_c$.
Dotted lines refers to HM as described for Fig.~\ref{thinpro}. 
To avoid overlap, a constant value of 1 magnitude has been
added to the surface brightness for the $\tau_V$=10 profile.
The right panel presents the difference in magnitude between each model
and the HM with the optical depth $\tau_V$.
All curves have been smoothed with a box of 10 pixels.}
\label{embpro} 
\end{figure*}

\begin{figure*}
\centerline{\psfig{figure=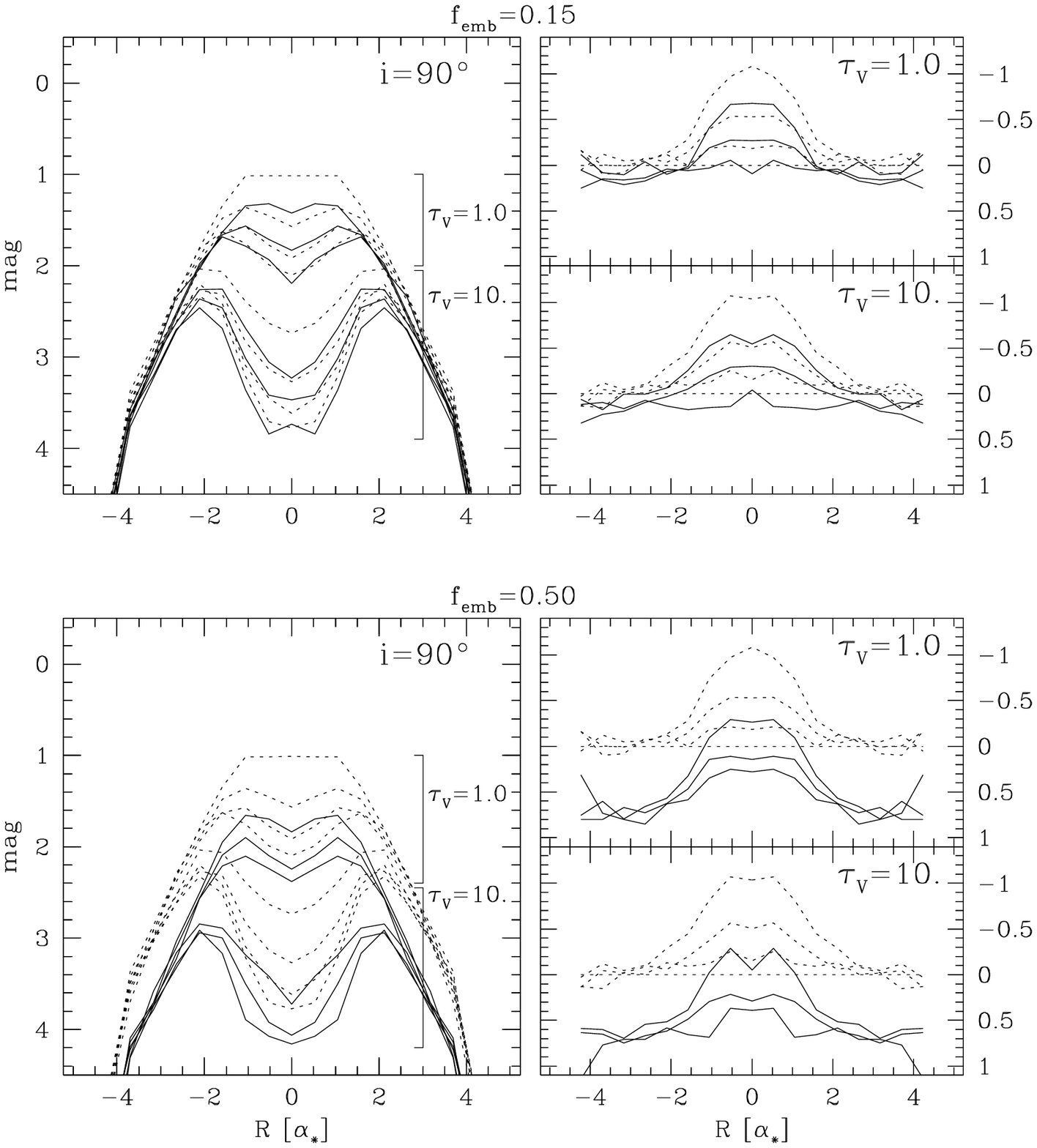,width=17.0cm}}
\caption{Edge-on minor axis profiles (left panels) for models with embedded 
sources.  Profiles are plotted for optical depths $\tau_V$=1, 10 and for the
three values of $f_c$ (solid lines). As in Fig.~\ref{thinpro}, brighter
profiles corresponds to higher values of $f_c$.
Dotted lines refers to HM as described for Fig.~\ref{thinpro}.
The right panel presents the difference in magnitude between each model
and the HM with the optical depth $\tau_V$.
}
\label{embmin} 
\end{figure*}

Face-on images, especially in the case for a higher fraction of embedded 
emission, clearly show a ring structure, due to the photons being emitted
in clumps distributed accordingly to the ring-like distribution of
molecular gas. When the models are seen edge-on, the ring is smoothed
out by the projection and it is clearly visible only for optically thin
cases and $f_{emb}$=0.50. In Fig.~\ref{embpro} we plot the major axis
profiles for the edge-on cases with $\tau_V=1 $ and 10. 
For $f_{emb}$=0.15 and for optically thick models with $f_{emb}$=0.50, the 
major axis profiles are similar to those of HM with different effective 
opacities.  It is interesting to note that the major axis profiles can be 
brighter than those of the HM with optical depth $\tau_V$, even when the
global opacities are larger. For example, in an edge-on model with 
$f_{emb}$=0.50, $f_c$=0.75 and $\tau_V=10$ only  20\% of the radiation 
escapes, while for the corresponding HM the fraction that escapes is 25\%. 
The profile along the major axis is brighter, being very similar to that of 
the HM with $\tau_V$=7.5. The brighter profile can be accounted for by the 
embedded emission, that is more concentrated in the plane than the smooth one.
While the CM is generally dimmer than the HM, the
major axis emission is higher: embedding concurs therefore with dust
clumping in reducing the contrast of the dust edge-on absorption lane.
This is clearly seen in the edge-on minor axis profiles in
Fig.~\ref{embmin}, especially for high optical depth and $f_{emb}$=0.50.

\section{Summary and Discussion}

We have presented in this paper the results of Monte Carlo radiative
transfer simulations through a two-phase clumpy dust distribution, for 
geometries typical of a spiral galaxy disk. The space occupied by the dust
distribution has been divided into a three-dimensional grid and each cell has
been assigned a clump or smooth medium status. The dimension of the
cells and the dust mass of a clump have been derived from the properties
of Galactic GMC. Clumps have been randomly distributed according to
the ring-like distribution of molecular gas observed in our Galaxy.
We have explored several values for the optical depth $\tau_V$
(i.e. the optical depth of a homogeneous dust distribution with the same 
mass) and for the fraction of total gas residing in clumps.

The main conclusions are the following.
As predicted by simple arguments and previous studies, a model with
a clumpy dust distribution (CM) suffers a reduced extinction. For the
parameters adopted in this paper, which emulate real dust distributions
in spiral disks, the reduction of the fraction of
absorbed energy is moderate, resulting in surface brightness
profiles that are always less than one magnitude brighter than the
corresponding homogeneous model (HM). The major differences between 
HMs and CMs are found for edge-on
inclinations. This is in contrast to the results presented by
Kuchinski et al.~\shortcite{ku98}. Using a model with spatially constant 
clump filling factor and with almost all (95\%) of the galactic 
dust locked in clumps, they find that the smallest differences occur in
edge-on cases. To ascertain the reasons for this discrepancy, we
have reproduced their experiment and confirmed their results. 
We have concluded that the disagreement depends on the different
parameters and distribution adopted for the clumps. 
This is  unfortunate, however, as it indicates a strong dependence of 
the observed brightness profiles on the detailed internal and spatial 
distribution properties of clumps which makes the interpretation of the 
data very difficult.

Since star formation occurs in high density clouds, it is logical to
assume that part of the stellar emission occurs within clumps. We have
therefore produced models where a fraction of the photons is emitted in
the high density cells. When this clumpy stellar distribution is 
considered, CMs are less transparent than for a clumpy dust distribution 
only. Depending on the fraction of gas in clumps and on the fraction of 
embedded stellar emission, galaxies can be even more opaque than predicted by
HMs with the same mass.

One of the major concerns about clumping is that its neglect will
produce a significant underestimate of the dust mass of a galaxy.
Xilouris et al.~\shortcite{xi99} analysed a sample of seven edge-on galaxies
fitting the surface brightness with a radiative transfer model.
They find a mean central face-on optical depth $\tau_V$=0.5. Comparing 
the total dust mass of each galaxy with the mass of gas, they derive a
gas-to-dust mass ratio of 360$\pm$60. The derived value is larger than the 
Galactic value by more than a factor of two \cite{so94}, but
closer than estimates based on FIR dust emission observed by IRAS 
\cite{de90}. From our results in Fig.~\ref{thinpro} and
Fig.~\ref{allmin}, minor and major axis profiles for CMs with
$\tau_V$=1 and $0.50<ff<0.75$ are similar to those for a HM
with $\tau_V=0.5$ (In the $\tau_V=1$ panels, a profile for a HM
with $\tau_V=0.5$ is shown between the two mentioned
CMs with a dotted line). Therefore, a CM with 
$\tau_V=1$ can have an edge-on appearance  very similar to an
HM with $\tau_V=0.5$. The mass of dust in the CM
would be twice the value derived for the HM and the gas-to-dust
mass ratio would be reduced accordingly to 180, a value close to the
canonical.  Similar results can be obtained using the $\tau_V=1$
profiles for models with embedded stellar emission.
We caution here that Xilouris et al.~\shortcite{xi99} find that dust disks 
have a larger radial scale lengths than the stellar while in the models of 
this paper both disks have the same radial scale lengths; this  may affect 
the numerical details of the exercise carried out in this paragraph, but
the qualitative result holds. 
On the contrary, a CM with a constant filling factor 
all over the galaxy  would resemble, when seen edge-on, a HM
with the same optical depth, and there would be no change in the
mass estimate. 

\vskip 0.7cm
\noindent {\em Acknowledgements.}
We wish to thank D. Galli for interesting discussion about the embedding
of stars in molecular clouds and the referee, A.
Witt, for useful comments that improved the presentation of the data in
the paper.

\end{document}